\documentclass[final,3p,times,authoryear]{elsarticle}

\usepackage{amssymb}
\usepackage{graphicx,amsmath}
\usepackage{ifthen}
\usepackage{xr}
\usepackage{multirow}
\usepackage{url}
\usepackage{hyperref}
\hypersetup{colorlinks=true,citecolor=blue}
\usepackage{breakurl}

\def\beq{\begin{equation}}
\def\eeq{\end{equation}}
\def\bea{\begin{eqnarray}}
\def\eea{\end{eqnarray}}

\biboptions{round,comma,sort&compress}

\journal{.}

\begin{document}

\begin{frontmatter}


\title{Complexity, economic science and possible economic benefits of climate change mitigation policy}

\author[4cmr]{J.-F. Mercure  \corref{cor1}}
\ead{jm801@cam.ac.uk}

\author[ce]{H. Pollitt}
\author[ce]{U. Chewpreecha}

\author[4cmr]{P. Salas}
\author[4cmr]{A. M. Foley}

\author[ou]{P. B. Holden}
\author[ou]{N. R. Edwards}

\address[4cmr]{Cambridge Centre for Climate Change Mitigation Research (4CMR), Department of Land Economy, University of Cambridge, 19 Silver Street, Cambridge, CB3 1EP, United Kingdom}

\address[ce]{Cambridge Econometrics Ltd, Covent Garden, Cambridge, CB1 2HT, UK}

\address[ou]{Environment, Earth and Ecosystems, Open University, Milton Keynes, UK}

\cortext[cor1]{Corresponding author: Jean-Fran\c{c}ois Mercure}
\fntext[fn1]{Tel: +44 (0) 1223337126, Fax: +44 (0) 1223337130}

\begin{abstract}

Conventional economic analysis of stringent climate change mitigation policy generally concludes various levels of economic slowdown as a result of substantial spending on low carbon technology. Equilibrium economics however could not explain or predict the current economic crisis, which is of financial nature. Meanwhile the economic impacts of climate policy find their source through investments for the diffusion of environmental innovations, in parts a financial problem. Here, we expose how results of economic analysis of climate change mitigation policy depend entirely on assumptions and theory concerning the finance of the diffusion of innovations, and that in many cases, results are simply re-iterations of model assumptions. We show that, while equilibrium economics always predict economic slowdown, methods using non-equilibrium approaches suggest the opposite could occur. We show that the solution to understanding the economic impacts of reducing greenhouse gas emissions lies with research on the dynamics of the financial sector interacting with innovation and technology developments, economic history providing powerful insights through important analogies with previous historical waves of innovation.

\end{abstract}

\begin{keyword}

Climate policy \sep Economic Theory \sep Complexity Science \sep  Technology and innovation

\end{keyword}

\end{frontmatter}


\section{Introduction}

The common perception of climate change mitigation in the economic perspective tends towards one of necessary future economic austerity \citep[e.g.][]{Nordhaus2010, IPCCAR5WGIIITS2, Edenhofer2010}, and this falls within a context of actual austerity policy in many nations due to the economic crisis \citep{Hourcade2013}. Many thus conclude that a world in recession cannot afford the low carbon technology necessary for avoiding dangerous climate change beyond~2$^{\circ}$. Consensus is now clear: current levels of emissions must be reduced by over 75\% by 2050, and to zero by 2100 and thereafter, in order to have a at least 50\% chance of not exceeding~2$^{\circ}$C \citep{IPCCAR5WGIIITS2,Meinshausen2009}. Policy for supporting the diffusion of low carbon innovations is, however, trailing behind with insufficient progress for meeting these critical targets. 

The economic perspective that the world cannot afford climate policy stems from general equilibrium theory in economics. Effectively, the profession of economist is, and has been for a long time, divided into two opposing clans; one which understands the economy as supply driven and in perpetual equilibrium\footnote{Neoclassical economics, general equilibrium theory.}, and all the other main schools of thought\footnote{This includes, but is not limited to, Keynesian, post-Keynesian, Marxian, post-Schumpeterian, Evolutionary and Complexity Economics.} that understand the economy as demand driven and out of equilibrium, featuring endogenously periods of prosperity and depression. In the equilibrium perspective, the economy uses economic resources (labour and capital, where real and financial capital are not distinguished) optimally at every point in time, and thus diverting resources, especially capital, for environmental objectives, necessarily takes away resources from all other uses, which then suffer from underinvestment. For that very reason, some supporters of the theory also use it to advocate least possible intervention in markets for obtaining socially optimal outcomes, and the positive equilibrium theory is thus not always distinct from a normative one. The balance of climate change mitigation policy generally tips towards reduced GDP, by an amount roughly equal to actual spending on low carbon technology.\footnote{In most Computable General Equilibrium (CGE) models of intermediate complexity \citep[e.g.][]{Waisman2012}, this is approximately true but depends on policy for revenue recycling. In a simple neoclassical model \citep[e.g.][]{Nordhaus2010,Tol2013,Stern2007}, this is by construction.} This explains why the problem of global mitigation is framed in the UNFCCC as one of burden-sharing. Meanwhile, in a non-equilibrium perspective, the outcome is generally different, and can tip both ways \citep[e.g.][]{Sijm2014, Lee2014, Barker2015}. In the climate change research area, practically all models are based on equilibrium theory, and thus results of the IPCC are predominantly oriented towards GDP reductions as a result of emissions reduction policy, such as for instance carbon pricing. 

As we show here, the crux of the matter is, however, that it is assumptions over the \emph{finance} of technology and innovation that is determinant of model outcomes, differentiating financial from real (e.g. production) capital. It also turns out that the same problem is at the root of the lack of predictive power of equilibrium economic theory to foresee the 2007 financial crash and current economic crisis \citep{Keen2011}. Here, we show that the choice of economic theory is determinant to results of any analysis of the economic impacts of climate policy. Using a non-equilibrium model, we explore the mechanisms involved in economic change related to emissions reduction policies, using the electricity sector as an example of finance-driven low-carbon technology investment. We calculate the environmental and economic impacts of global electricity sector decarbonisation in 21 regions of the world, and show that in the non-equilibrium perspective without limits to finance, economic benefits can emerge from policy supporting the diffusion of low carbon technology, through investment and job creation, in other words, the use of unemployed resources in the economy. It is thus the amount of finance available for low carbon ventures across the world that is the unknown parameter.

In the economic exercise presented here, two interacting non-equilibrium methodologies are used: investor level technology selection and diffusion is specifically represented in a bottom-up technology model, which enables to simulate the evolution of the electricity sector without an optimisation procedure, embodied in the model FTT:Power \citep[see appendix \ref{app:A} for model descriptions, and ][for more details]{Mercure2012, Mercure2013b}. This is integrated with dynamical feedbacks in a highly disaggregated non-equilibrium macroeconometric model used to simulate the evolution of the global economy and greenhouse gas emissions, E3MG \citep[variant of E3ME, \burlalt{http://www.e3me.com}{www.e3me.com}, see ][]{Barker2010,Barker2006,Kohler2006}. The emissions are fed to emulators \citep{Holden2010} of the carbon cycle \citep[GENIE1-em][]{Holden2013} and of the climate system \citep[PLASIM-ENTS-em][]{Holden2013b} in order to determine the environmental impacts of electricity policy choices. Following our previous technology-focussed work \citep{Mercure2014}, we  demonstrate an ability, with this combination of non-equilibrium simulations, of modelling the environmental impacts of specific individual and groups of policy instruments through the global energy sector and the economy.

These results, however, essentially depend on our assumptions about the current ability to finance low carbon developments. In the light of results of the recent IPCC report, we conclude that it is crucial that a debate opens concerning assumptions and theory on the economics and finance of low-carbon technology before advising policy-makers on climate policy, and that more research must be carried out on the complex mechanisms involved in the financial sector, which, in the right conditions, could channel resources to enable simultaneous emissions reductions and enhanced global economic development and growth: \emph{green growth}.

\section{The choice of economic theory and its impact to conceptualising environmental policy-making}

Emissions reductions necessarily involve the diffusion of low carbon innovations.\footnote{E.g. electric cars, renewable energy systems, carbon capture and storage.} As Brian Arthur famously demonstrated, the diffusion of technology, when featuring increasing returns to adoption,\footnote{Increasing probability of adoption with increasing adoption.} lead to more than a single equilibrium \citep{Arthur1989}. In fact, an economy that features increasing returns (i.e. positive feedbacks) generally does not have a single equilibrium but instead can feature several, limit cycles or even chaotic behaviour typical of dynamical systems \citep{Arthur2014}. For instance, stochastic systems of interacting financial investors in the stock exchange feature emergent complex structures that can significantly differ in shape and level of complexity \citep{Palmer1994, Arthur2014}. In such cases, it has become clear to many that economic science could benefit from general insights derived from Complexity Theory \citep{Anderson1972, Arrow1995, Anderson1988}, where structures and \emph{ordering principles} emerge from the interaction between system components, an endeavour that could be called Complexity Economics \citep{Arthur2014}. 

It has been long known that the major component of economic (or productivity) growth, is attributable to technological change \citep{Solow1957}. The diffusion of new technology necessarily involves innovation, the economic impacts of which were first described a century ago by \cite{Schumpeter1934} and after by his living legacy of Evolutionary Economics \citep[e.g.][]{Nelson1982,Andersen1994, Safarzynska2010}. In Schumpeter, the calm of equilibrium is incessantly disturbed by credit financed entrepreneurial innovative activity, which continuously replaces old socio-technical regimes by new ones (`creative destruction'), thereby continuously increasing productivity, leading to economic development. Many sectors of innovation being interlinked by cross-sectoral spillovers, innovation clusters in waves, leading to business cycles at different timescales, producing booms, recessions and depressions \citep{Schumpeter1939}, corroborated by facts well documented in economic history \citep{Freeman2001}. Entrepreneurs generally requiring finance in order to successfully develop their ideas, historical waves of clustered innovation have taken place in tandem with financial sector developments, which have typically led to expectations, bubbles and financial meltdowns \citep{Perez2002}.\footnote{For instance the 1793-1797 canal mania, the 1836-1847 railway mania, the 1929 Great Depression and the 2000 information technology bubble.} Increasing private debt indirectly contributing to aggregate demand \citep{Schumpeter1934,Keen2011}, economic growth has historically been closely linked to economic development \emph{and} cycles, tied to financial expectations and the ability of agents to repay debt. 

Unfortunately leaving innovation and technological progress outside the focus of analysis,\footnote{Sometimes in the form of an independent historical trend of technological change, with the notable exception of models with induced technological change \citep{IMCP2006}.} equilibrium theory excludes the existence of increasing returns in the economy,\footnote{Such as tied to learning-by-doing, imitation by consumers, contagion-type diffusion of ideas, practices and technologies and expectations.} in order to ensure a unique equilibrium. Meanwhile, representing diverse individuals and households with a single `representative agent' enables to solve the intractable problem of aggregating the utility of an ensemble of otherwise interacting diverse agents with a distribution of income, which would also lead to several solutions to utility maximisation.\footnote{The only way utility-maximisation can be ensured to have a unique maximum is if agents are identical. Furthermore, the only way that the maximum of the aggregate utility function can correspond to the maximum of the utility of every agent is if they are identical. If they are not identical (e.g. there is a distribution of income), some agents are altruistic and utility maximisation has several solutions, none of which is socially optimal.} 

The key issue there is that this approach leaves out information which may actually be crucial. Indeed, both increasing returns \citep{Arthur1989} and the diversity of agents \citep{Mercure2013b} lead to multiple equilibria, and thus to \emph{path-dependence} where events in the present narrow down the range of possible future events, emphasising the meaning of the passage of time in the theory. In a path-dependent (and thus non-equilibrium) perspective, economic scenarios behave similarly to climatology, where the uncertainty of predictions, tied to the uncertainty in starting parameters, increases exponentially with forecasting time. Path-dependent scenarios can feature limit cycles, attractors and chaos typical of dynamical systems. They are simulated by solving systems of difference/differential equations as opposed to finding an equilibrium using optimisation/linear programming techniques, leading to a sheer methodological divide between the schools of thought. Finally, in equilibrium theory, since the representative agent has perfect foresight (i.e. rational expectations), he never defaults on his debts, and in a theory without uncertainty or risk on investment, where the future is known by all agents, a financial sector has no meaning.

The theories of Marx, Keynes, Schumpeter and Minsky, as well as their respective modern followers, have in common that they do not involve an economic equilibrium. Quoting Steve \cite{Keen2011}, one of the few economists who predicted the 2007 financial crisis, 
\begin{quote}
\small
The non-Say/Walras's Law [non-equilibrium] vision of the economy shared by Marx, Schumpeter, Keynes and Minsky thus accords with the manifest instability of the economy, whereas Walras's Law [equilibrium theory] asserts that, despite appearances to the contrary, the macroeconomy is really stable. 

At the same time, this potential for instability is also a necessary aspect for the potential for growth. [...] The neoclassical obsession with equilibrium is therefore a hindrance to understanding the forces that enable the economy to grow, when growth has always been a fundamental aspect of capitalism. \citep[][p. 224]{Keen2011}
\end{quote} 

Indeed, it is perhaps the ability of the economy to develop through the diffusion of disrupting innovations, and the speculation that accompanies it, that is not clearly represented in equilibrium theory. This happens through the theoretical restriction in the number of possible dynamics to solely those that lead to a unique equilibrium, in particular those related to finance and technology diffusion.

This has clear implications to conceptualising environmental policy-making. In equilibrium theory, the equilibrium is understood as delivering socially optimal outcomes, and thus leads to important relationships to ethics. In a theory where any intervention for environmental protection invariably leads to reduction of consumption and welfare, particularly to the poor, it is understandable that immediate questions on ethics and social justice emerge tied to the theory, and dominate the debate \cite[e.g.][Ch. 2]{IPCCAR5WGIIITS2}. This is much more subtle in a non-equilibrium perspective, since as we show below, it is not immediately obvious what the impacts to consumption are of technology and emissions reductions policy. Instead, it strongly depends on the form and type of policies used, and on \emph{mechanisms for wealth redistribution} across industrial sectors (e.g. carbon pricing), economic function (e.g. finance mechanisms for entrepreneurs) and income groups (e.g. income tax). Meanwhile, it is the diffusion of innovations that create productivity growth and wealth, but it also leads to speculation, bubbles, crashes and recessions (e.g. as happened with the dotcom bubble in 2000).

Theoretical work in non-equilibrium complexity economics however does not yet provide a complete explanation of the relationship between support for technological change and economic development. As we show below, supporting technological change could generate employment and wealth, but this assumes the availability of financial and production capital to support investment intensive developments as required in the decarbonisation of the global economy. In a pure equilibrium model, no `spare' finance and production capital exists to support low carbon developments: they \emph{crowd out} investments elsewhere. In a non-equilibrium macro model of the real economy, the output gap can be estimated, enabling to assess whether sufficient spare production capital exists to carry out such developments (without crossing Keynes' \emph{inflationary gap}). However, without a representation of the financial sector, current theory does not yet determine whether lenders would be willing to finance major low carbon investments. In a credit starved global economic context pervaded by uncertainty, this is the true constraint to concurrent emissions reductions and economic development, which is currently unknown.

\section{A non-equilibrium technology diffusion perspective}

\begin{figure*}[t]
		\begin{center}
			\includegraphics[width=1\columnwidth]{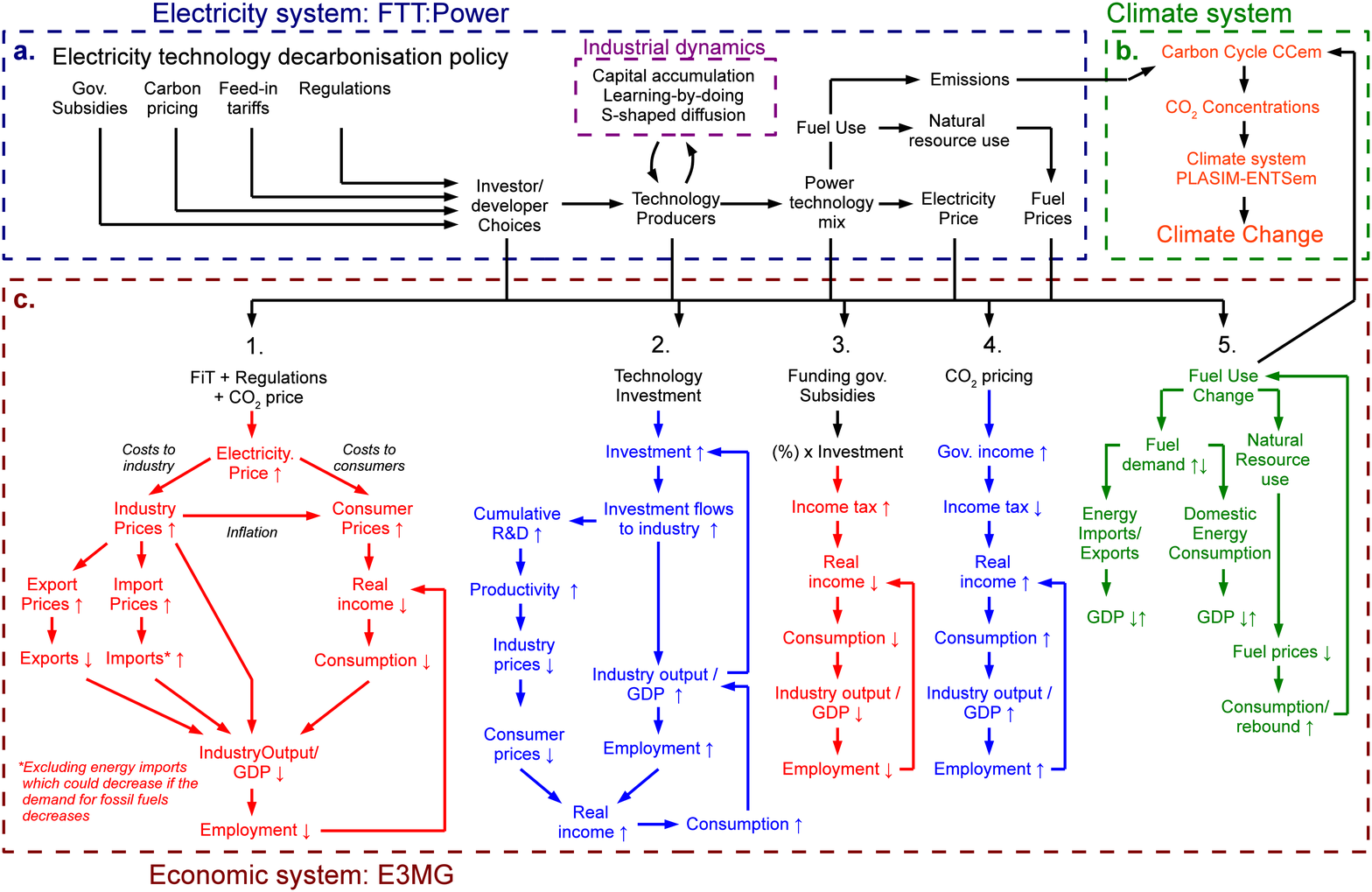}
		\end{center}
	\caption{Map of all economic feedbacks operating during the process of decarbonisation of the power sector in the combination of models FTT:Power (\emph{a.}), GENIEem-PLASIM-ENTSem (\emph{b.}) and E3MG (\emph{c.}). See appendix \ref{app:A} for more details on the models. Red areas indicate processes contributing to economic slowdown, while blue areas indicate positive feedbacks. The green area indicates fuel import/export feedbacks beneficial to some regions while detrimental to others.}
	\label{fig:Figure1}
\end{figure*}

Figure~\ref{fig:Figure1} lays out the conceptual components of FTT:Power-E3MG translating climate policy into macroeconomic impacts in 21 regions of the world. Technology diffusion corresponds to investor choices filtering available options based on accessible, uncertain and incomplete information (bounded rationality), influenced by policy, generating a form of natural market selection (panel~\emph{a}). We have shown elsewhere that resulting competitive industrial dynamics with technology producers constrained by technology lifetimes and industry growth rates produce technology population dynamics similar in form to those occurring with species in ecosystems competing for resources \citep{Mercure2013b}, which can be derived from first principles. This includes logistic and Lotka-Volterra diffusion patterns and learning-by-doing \citep{Mercure2012, Saviotti1995, Metcalfe2004, Hofbauer1998}, a proposition well substantiated by the empirical literature \citep{Grubler1998, Grubler1999, Wilson2012b, Wilson2009, Mansfield1961, Fisher1971, Sharif1976, Weiss2012}. In contrast to the cost-optimisation approach, this theoretical basis is appropriate to model path-dependent profiles of technological change, their timescales, and well known technology lock-ins at odds with minimisation of cost. In this perspective, mitigation action corresponds to channelling investor choices and capital flows towards lower emission technology systems. Resulting emissions changes impact the level of future climate change, represented here using emulators of the carbon cycle, GENIEem \citep{Holden2013}, and of the climate system, PLASIM-ENTSem \citep[][panel~\emph{b}, for brief model descriptions see appendix \ref{app:A}]{Holden2013b}.

In the economic sphere, (panel~\emph{c}, showing with blue (red) positive (negative) impacts of climate policy onto economic growth), abatement costs can be decomposed into real economic components: changes in the price of electricity and fuels, enhanced investments in technology, additional climate policy-related government spending and income. Here, specific climate policy instruments are simulated: carbon pricing (CO$_2$P), technology subsidies (TSs), feed-in tariffs (FiTs) and direct technology regulations (REGs), influencing investor choices. Using multiple policy instruments in a path-dependent model with slow timescales of industrial adjustment means that the carbon price is not a single valued function of emissions reductions,\footnote{The emissions reduction effectiveness of a carbon price depends, amongst many things, on context, synergies between policies \citep{Mercure2014}, its rate of change, and complex cross-sectoral interactions.} and thus cannot be equated to a social cost of carbon to derive normative conclusions, as commonly done in cost-benefit analysis integrated assessment models \citep{Nordhaus2010,Stern2007,Tol2013}.

Their impact goes much further: while CO$_2$P is generally passed on to consumers through a raise of the price of electricity, FiTs generate further price rises that ensure access to the grid for costly electricity generation methods. REGs over what can be built or not alter electricity production costs, also reflected in the price. While many economic sectors can avoid part of these price rises by reducing their consumption, they nevertheless all require electricity, thus these changes propagate throughout the economy, resulting in higher industrial and consumer price levels, and thus inflation. Household real income is reduced followed by consumption which, including a Keynesian multiplier effect (resulting from cross-sectoral linkages), slows down economic activity and impacts employment, creating a vicious cycle. These effects are temporary, however, since higher  energy technology investments trigger enhanced energy efficiency (see \emph{Appendix A}), which enable to maintain levels of industrial and household activity with lower electricity consumption and thus lower costs.

\begin{figure*}[t]
		\begin{center}
			\includegraphics[width=1\columnwidth]{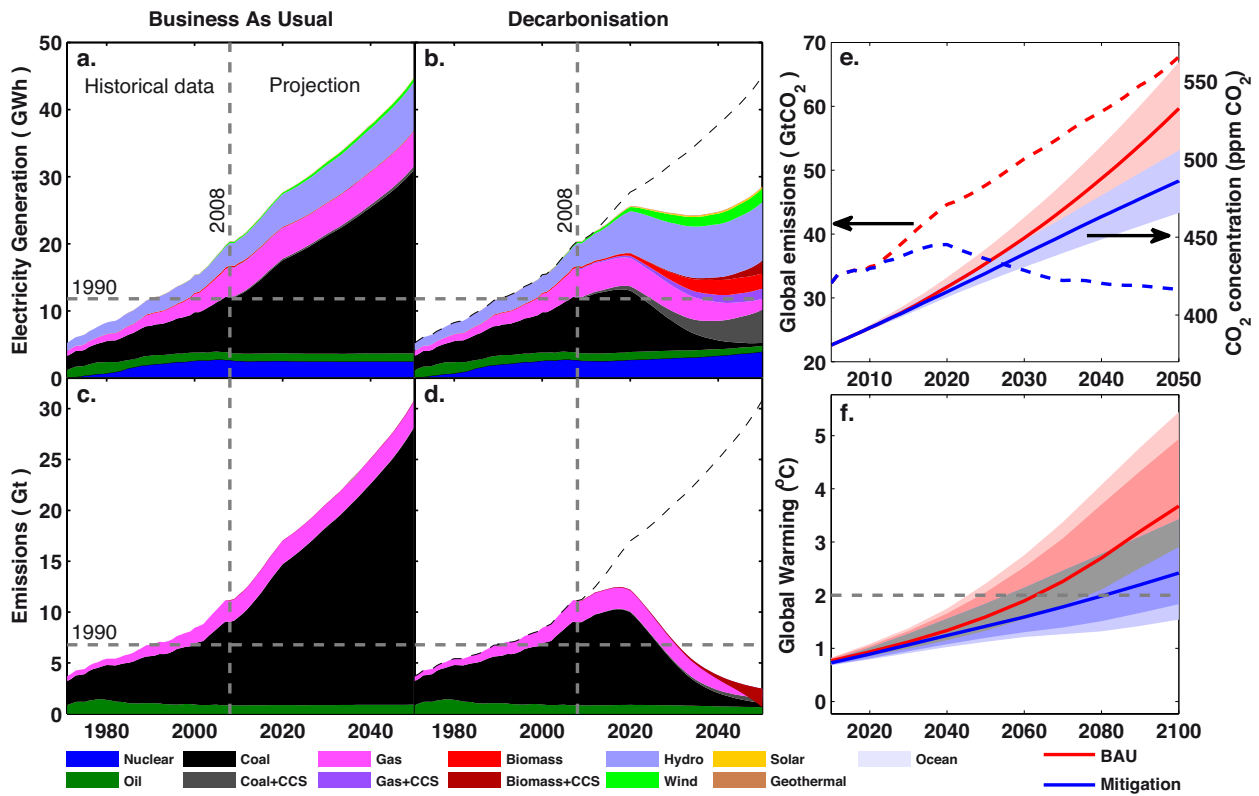}	
		\end{center}
	\caption{Global summary of (\emph{a} and \emph{c}) a baseline scenario assuming current policies and (\emph{b} and \emph{d}) a mitigation scenario assuming coordinated mitigation action across regions of the world. The top panels show electricity generation by technology, while the lower panels present the associated CO$_2$ emissions. Note that the red area of emissions by Biomass+CCS is a negative contribution of sequestrated emissions, \emph{reducing} global emissions below the top curve. The dashed vertical lines delimitate historical IEA World Energy Balances data from \emph{FTT:Power-E3MG} projections, while the horizontal dashed lines indicate the 1990 emissions level. Resulting global emissions are given in panel~\emph{e} overlaid with corresponding CO$_2$ concentrations, where colour patches delimitate the 95\% confidence interval from the climate model. Global warming temperatures are given in panel~\emph{f.}}
	\label{fig:Figure2}
\end{figure*}

Low-carbon technologies are more investment intensive than traditional (fossil) systems \citep{IEAProjCosts}. Propagating throughout industry with a Keynesian multiplier effect (cross-sectoral spillovers), climate policy induced enhanced investment produces increased industrial activity, and in the short term this generates employment and real income that counteracts the impact of higher electricity prices. We show below that in our mitigation scenario, the two effects roughly cancel out. This also incentivises a virtuous cycle of industrial capacity expansions and R\&D investment-driven productivity growth that contribute to economic growth in the long run, the source of endogenous growth in E3MG.

CO$_2$P and TSs are instruments that are used to bridge the difference in costs between inexpensive but polluting technology and new low carbon systems \citep[bridging the technology `valley of death' ][]{Murphy2003,Grubb2014}. In scenarios where CO$_2$P and TSs are chosen in such a way, in combination with other policy instruments, that decarbonisation takes place, this can generate significant income (CO$_2$P) or costs (TSs) to governments. We show below that the carbon pricing income is much larger than the subsidy costs, and in a scenario where government finances are kept revenue neutral (i.e. this income is not used to repay government debt), this means a potentially substantial redistribution of funds to other economic agents, households in this particular case, which is done by lowering income tax. This generates additional household consumption and increases GDP.  

\section{Global power sector decarbonisation}

The decarbonisation of the global electricity sector was simulated up to 2050 with FTT:Power-E3MG using various combinations and strengths of the four different types of energy policy instruments. The resulting emissions were fed to GENIEem in order to obtain concentrations and to PLASIM-ENTSem for climate impacts. 10 such scenarios were created, explored in detail previously elsewhere \citep{Mercure2014}, with scenario data available online at \burlalt{http://www.4cmr.group.cam.ac.uk/research/FTT/fttviewer}{www.4cmr.group.cam.ac.uk/research/FTT/fttviewer}. We show there that the impacts of coordinated sets of policies do not correspond to the sum of the impacts of individual policies, and thus that strong \emph{synergies} exist between instruments. A detailed policy scheme in every E3MG-FTT region is built that in total decarbonises the global electricity sector by 90\% below its 1990 emissions level before 2050, summarised in figure~\ref{fig:Figure2}. The baseline scenario (panel~\emph{a}) is characterised by a strong fossil technology lock-in, which is difficult to break even with high CO$_2$P. While CO$_2$P is not sufficient, it is nevertheless necessary  to break the lock-in. The combined use of CO$_2$P, TSs, FiTs and REGs (See data in appendix B, fig.~\ref{fig:Figure2aSuppl}.) leads to fast investor uptake of low carbon technologies and power sector decarbonisation, characterised by increasing technology diversity and fast phasing out of fossil fuel power technologies as they come to the end of their lifetimes \citep[fig.~\ref{fig:Figure2},~panel~\emph{b}, see ][]{Mercure2014}. 

As shown in panels~\emph{c} and~\emph{e}, total emissions in the baseline reach 65~GtCO$_2$, cumulative emissions for the time span 2000-2050 of 2321 GtCO$_2$. and a concentration of $533\pm30$~ppm in 2050, very likely to lead to 4$^{\circ}$C of global warming or more in 2100 (panel~\emph{f}). Meanwhile, the decarbonisation scenario emissions reach 31~GtCO$_2$, cumulative emissions for the time span 2000-2050 of 1603 GtCO$_2$ and a concentration of $485\pm20$~ppm. We observe that even if the power sector decarbonises by 90\%, it is likely not enough to maintain global average temperatures below a median average temperature rise value of 2$^{\circ}$C or less if other sectors of anthropogenic emissions do not change significantly their technologies (panels~\emph{d},~\emph{e},~\emph{f}). Consistent cross-sectoral climate policies are required in order to curb total emissions below 20~Gt or less in order to maintain temperatures below 2$^\circ$C beyond 2100 \citep{Meinshausen2009}. For example, further reductions of 12-15~Gt could be achieved in the transport, household and industrial sectors, likely to bring down concentrations below 450~ppm, generally considered a safe level.

It is to be noted, following the dashed lines of panels~\emph{b} and~\emph{d} referring to the baseline, that a large contribution to emissions reductions stem from reductions of electricity consumption in a context of higher electricity prices. Depending on the nature of the redistribution of government income, this opens a discussion over potentially increasing energy poverty as a result of climate policy: climate change mitigation offers an opportunity and framework of analysis to address the planning of bringing energy access for all in under-developed countries \citep{GEA2012ch17}.

\section{Economic impacts of climate policy}

\begin{figure*}[t]
		\begin{center}
			\includegraphics[width=1\columnwidth]{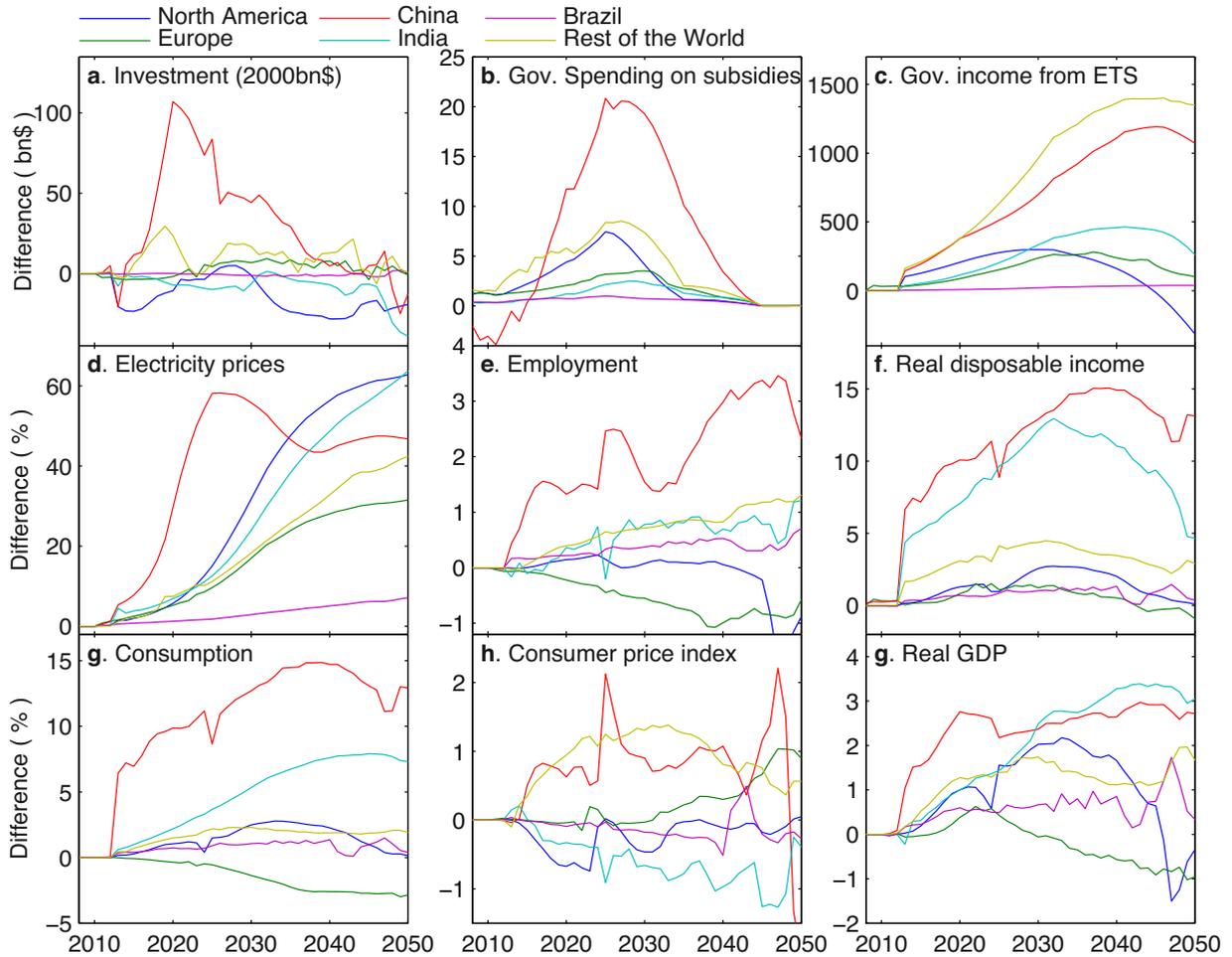}
		\end{center}
	\caption{Summary of economic changes due to climate policy resulting in the decarbonisation of the electricity sector. Values are expressed as changes from the baseline scenario in six aggregate region, in billions of constant 2000 US dollars (\emph{a}), in nominal US dollars (\emph{b,c}) or percent (\emph{d} to \emph{h}).}
	\label{fig:Figure3}
\end{figure*}

Figure~\ref{fig:Figure3} presents a summary of the economic changes occurring in E3MG following climate policy and decarbonisation, as a result of the sum of the feedbacks of fig.~\ref{fig:Figure2}. The sectoral transformation leads to enhanced investment in electricity technology (\emph{a}), TSs to government spending (\emph{b}) and CO$_2$P to government income (emissions $\times$ CO$_2$P,~\emph{c}), the latter larger than spending. CO$_2$P, FiTs and REGs lead to significant increases in electricity prices (\emph{d}), responsible for the major decrease in electricity consumption of fig.~\ref{fig:Figure1}. Enhanced investment and the redistribution of government income from CO$_2$P, after accounting for TSs, lead to additional employment (\emph{e}) and increases in household real disposable income (\emph{f}) and consumption (\emph{g}) albeit higher price levels (\emph{h}). GDP \emph{increases} in all world regions as a result of electricity sector decarbonisation (\emph{i}).

Further computations (See data in Appendix B, fig.~\ref{fig:Figure1Suppl}) show that a significant part of this positive push on GDP stems from the redistribution of CO$_2$P income back to households, which is spent again, a form of energy tax recycling \citep{Barker2010}. In a non-neutral revenue scenario without recycling, we find that global GDP remains mostly unchanged with mitigation, indicating that the price and investment feedbacks approximately cancel out. This also contrasts with standard equilibrium analyses, where without revenue recycling, costs to GDP would be substantial. We emphasise that this balance between higher investment and higher electricity prices are due to sectoral increasing returns to investment: investment driven productivity growth spilling across sectors despite higher prices. Thus the further decarbonisation of other sectors of emissions (transport, industry) also appears achievable without significant macroeconomic impacts, so long as finance resources are assumed available, as we do in this exercise.

A caveat arises here: our assumption that the diffusion of new power technology raises the electricity price in order to be economical inherently involves the electricity sector borrowing to finance heavy investment intensive technology (fig.~\ref{fig:Figure3} \emph{a.}).\footnote{It doesn't matter if money is borrowed from banks or if utility companies self-finance, as the same process takes place involving uncertainty over future profitability.} This is naturally paid back by passing the increased costs to consumers over the lifetime of the new generators. Hence, these scenarios involve taking resources from the financial sector, which is spent immediately, to be returned gradually over a long period of time (which, in some cases, spans beyond the end point of the simulations). These very resources are at the source of the investment-led economic activity and productivity growth that counteracts the electricity price effect, particularly in the early parts of the simulations. For example, GDP grows in Europe during the initial period where massive investment in renewables take place, but declines below the baseline afterwards when this is completed and consumers pay back the loans of the utility companies.

Therefore it is clear that the financial sector drives aggregate demand and economic growth in this policy-led diffusion process. What is unknown here is the willingness of financial institutions to lend this amount of funds. In equilibrium economics, the amount of capital, real or financial, is constant, and therefore no additional funds would be available unless they are taken out from another sector. Here, we make a distinction between production and financial capital, both of which are not constant. The amount of financial capital is determined by the willingness of financial institutions to fund entrepreneurial ventures, based on their expected return and risk of default, not by the amount of money available in the economy. A recent review article issued by the Bank of England \citep{BoE2014}, as well as the standard post-Keynesian literature \cite[see][]{Keen2011},\footnote{Note that \cite{Schumpeter1934, Schumpeter1939} argued this was the case since at least as far back as 1911, as well as Keynes, who argued it in his General Theory \cite{Keynes1936}.} suggest that this is indeed how lending takes place: banks create money. In this perspective, there is no limit to financial resources beyond the amount of ventures that banks find worthwhile funding. A fluctuating amount of finance in the economy is \emph{also} what enables speculation bubbles to form and burst, leading to the possibility of economic crises.

\section{Conclusion: Economic theory and the climate policy dialogue}

Policy for the diffusion of low carbon innovations presents important opportunities for industry \citep[e.g.][]{CBI2014} and job creation \citep[e.g.][]{UKERC2014}. The interaction between innovation sectors, the diffusion of new environmental technologies and economic development arising from these activities is poorly understood. There is an apparent contradiction between observations of new activity arising in highly successful low carbon ventures (e.g. Tesla electric and Toyota hybrid cars, wind turbines, solar photovoltaic), and the common perception, stemming from the current mainstream equilibrium economic theoretical paradigm, that the use of low cost fossil energy and technologies is indispensable for sustained future growth, and therefore that a choice has to be made between growth and environmental protection. This dichotomy in the public consciousness is an expression of the current crisis in theoretical economics, applied to environmental problems: depending on theory, we have contrasting understanding and results, and this leads to lack of consensus when advising policy-making. 

We have first shown here that a significant amount of investment is required to decarbonise the electricity sector in order to increase the chance of avoiding global warming beyond 2$^{\circ}$C, and that electricity decarbonisation is insufficient to avoid such dangerous warming without decarbonising other sectors as well. With these investments alone, however, we have shown that in a highly detailed, sectoral and regional non-equilibrium model of the real global macroeconomy without limits to finance, policy supporting the diffusion of low carbon technology can lead, in particular contexts of revenue recycling, to simultaneous enhanced economic growth and income, and rapid emissions reductions. Meanwhile, we know from standard climate change mitigation economics literature \citep[e.g.][]{IPCCAR5WGIIITS2,Edenhofer2010,Stern2007,Nordhaus2010} that in an economic theory with constant amounts of economic resources, spending on low carbon technology \emph{always} leads to economic losses. A recent statement from the Bank of England was issued expressing that this assumption over financial resources, and therefore on the money supply, does not quite represent reality correctly \citep{BoE2014}, re-iterating the general result of post-Keynesian theory \citep{Keen2011} in which the money supply is endogenous.

We conclude by emphasising that it is the availability of finance to support the diffusion of low carbon innovations that really determines the outcome, a field not completely well understood yet, and not discussed at all in climate change mitigation research. This is intimately related to our current lack of consensus over the dynamics underlying the current debt-related economic crisis, where banks are reluctant to finance ventures that they perceive as excessively risky. We thus conclude in this work that an improved understanding of the dynamics driving the current crisis will unlock crucial insight to understanding the economic impacts of climate change policy (or conversely), and that low-carbon technology policy could potentially help, rather than hinder, to promote new economic activity and employment in the form of investments in R\&D and technology deployment, driving productivity growth, in a global economy increasingly led by its knowledge base.

\appendix

\section{Model and assumption details\label{app:A}}

\textbf{Innovation/Diffusion in FTT:Power.} The diffusion theory underlying technology substitutions in FTT:Power is described in \cite{Mercure2013b, Mercure2012, Mercure2014}, in which the Lotka-Volterra equation of competing population dynamics is derived from demographic principles applied to technology, describing the associated industrial dynamics. It tracks the birth, ageing and scrapping of technology units in time, enabling a realistic representation of non-linear rates of substitution respecting technology lifetimes and competition. This is quite important for the power sector, in which technologies have very long and varied lifetimes. The non-linear differences equation calculating these changes independently in each of the 21 world regions is 
\beq
\Delta S_i = \sum_j S_i S_j \left(A_{ij} F_{ij} G_{ij}-A_{ji} F_{ji} G_{ji} \right){1 \over \overline{\tau}} \Delta t,
\eeq
where the $S_i$ represent the capacity shares of technologies $i$ in a particular world region. This equation effectively represents \emph{flows} of \emph{market shares} between technology options. The matrix $F_{ij}$ represents probabilistic choices by diverse investors derived from a binary logit choice model \citep{Mercure2013b}, and parameterised by market data \citep{IEAProjCosts, Mercure2012} while the matrix $A_{ij}$ provide the frequencies of substitutions between all possible pairs of technologies. The matrix $G_{ij}$ maintains the system within technical constraints of operation (grid stability). $\overline{\tau}$ is an overall time scaling constant. Technology choices $F_{ij}$ are influenced by technology costs, which follow learning curves:
\beq
C_i = C_{i,0} \left({W_i \over W_{i,0}}\right)^{-b},
\eeq
where $C_i$ is the investment cost, $W_i$ is the cumulative capacity of a technology (including previous decommissions), $C_{i,0}$ and $W_{i,0}$ are values at the start of the simulation and $b$ is the learning exponent, related to the learning rate \citep{Mercure2012}. FTT:Power features 24 power technology categories, 21 regions and 13 types of primary natural resources. Projected costs of non-renewable fuels (oil, coal and gas) are done based on existing reserves and resources according to our model of commodity price dynamics \citep{MercureSalas2013b}.

\vspace{8pt}

\textbf{Macroeconometrics in E3MG.} E3MG \citep[variant of E3ME, see ][]{E3MEManual} is formed of 33 econometric relationships (energy demand, industrial investment, prices, hours worked and employment, imports, exports, etc). The magnitudes of these relationships are evaluated using multi-variate regressions onto historical data since 1970, currently using 43 industrial and investment sectors, 28 consumption categories, 21 World regions, 22 types of fuel users and 12 types of fuels. These relationships are used to project into the future the complete set of variables of the model (in each sector, region, category etc, generating thousands of equations), which includes economic output, investment, prices, consumption, employment, disposable income, imports, exports and CO$_2$ emissions that stem from fuel combustion. Meanwhile, sectors are connected to one another using input-output tables expressing intra-industry flows of goods and investments, resulting in typical Keynesian multiplier effects. The model does not assume optimal growth, economic equilibrium, full employment or use production functions \citep{Barker2010}. It is demand led and supply constrained.

E3MG features endogenous growth, with sectors interrelated using dynamic input-output tables \citep{E3MEManual}, but does not optimise economic resources: it is a simulation model. This is obtained following Kaldor's theory of cumulative causation \citep{Kaldor1957, Lee1990,Loschel2002}, where Technology Progress Indicators (TPIs) are created by cumulating past investments and R\&D spending using a relative time-weighting,\footnote{$\tau_1$ plays the role of knowledge depreciation while $\tau_2$ is a weighting parameter for R\&D.} 
\beq
TPI(t) \propto \sum_{a = 0}^\infty e^{-a \tau_1} \ln \left(I(t-a) + \tau_2 R\&D(t-a)\right).
\label{eq:TPI}
\eeq
Such TPIs are used in the regressions for industrial prices, international trade and employment. Lower prices incentivises higher consumption, and thus industrial investment and R\&D expenditures for production capacity expansions, which lead to lower prices and so on, producing a self-reinforcing cycle of \emph{cumulative causation of knowledge accumulation}, consistent with Schumpeter's theory of economic development through the intervention of the entrepreneur. Including accumulated investment and R\&D makes E3MG non-linear, path-dependent and hysteretic, and thus far from equilibrium.

Electricity demand is modelled in E3MG using an econometric equation that incorporates a contribution from spillovers from investment and R\&D spending in other sectors \citep{Forssell2000,E3MEManual}. Since new investments tend to involve technologies with higher energy efficiencies and because the turnover of capital does not allow return to old technology, here the TPI is formed by cumulating positive increases in investment and R\&D, which thus cannot decrease.\footnote{According to the study performed by \cite{Forssell2000}, this describes historical data better than the TPI given in eq.~\ref{eq:TPI}.} The energy demand equation takes the form (in log-linear scaling) 
\beq
\Delta \log X_{ik} = \beta^0_{ik} + \beta^1_{ik}\Delta \log Y_{ik} + \beta^2_{ik} \Delta \log P_{ik} + \beta^3_{ik} \Delta \log TPI_{ik} + \epsilon_{ik},
\eeq
where for fuel $i$ and region $k$, $X_{ik}$ is the fuel demand, $Y_{ik}$ represents sectoral output, $P_{ik}$ relative prices. As shown in fig.~\ref{fig:Figure1}, economic feedbacks between FTT:Power and E3MG occur with four quantities: electricity prices, fuel use, power technology investments and tax revenue recycling.

\vspace{8pt}

\textbf{Emulating the climate system.} GENIEem: The carbon cycle is represented in this work by an emulator \citep{Holden2010} of the Grid Enabled Integrated Earth systems model (GENIE-1) Earth System Model, a simulation of the climate, ocean circulation and sea-ice, together with the terrestrial, oceanic, weathering and sedimentary components of the carbon cycle \citep{Holden2013}. GENIEem is designed to be a more sophisticated and computationally faster alternative to simplified climate-carbon cycle models. GENIEem transforms E3MG-FTT emissions into atmospheric concentrations with an analysis of uncertainty performed using the ensembles method representing large numbers of runs of the original model.

PLASIM-ENTSem: The climate system is represented by an emulator of the Planet Simulator (PLASIM) climate, coupled to the Efficient Numerical Terrestrial Scheme (ENTS) \citep{Holden2013b}. PLASIM-ENTSem also uses the ensembles method for evaluating uncertainty, and is driven here by atmospheric concentrations calculated by GENIEem.  Therefore a second layer of uncertainty emerges in addition to that of the carbon cycle. The combination of both emulators in the context of this work, with combined uncertainty analysis, is described further in \cite{Mercure2014}.

Warming is in practice affected by the concentration of other `non-CO$_2$' GHGs such as CH$_4$ and N$_2$O. However, we use the same values for non-CO$_2$ forcing, taken from the RCP8.5, in both scenarios, owing to a lack of capacity for modelling emissions unrelated to fuel consumption (e.g. agriculture and forestry). E3MG-FTT projects reductions in fuel use-related CH$_4$ and N$_2$O emissions of around 10-15\% in the mitigation scenario, producing a small reduction in forcing of roughly 0.1~W/m$^2$ in comparison to the total forcing of 8.4~W/m$^2$ in the baseline and 4.3~W/m$^2$ in the mitigation scenario. We neglected this change of order 1-2\%.

\vspace{8pt}

\begin{figure}[h!]
		\begin{center}
			\includegraphics[width=1\columnwidth]{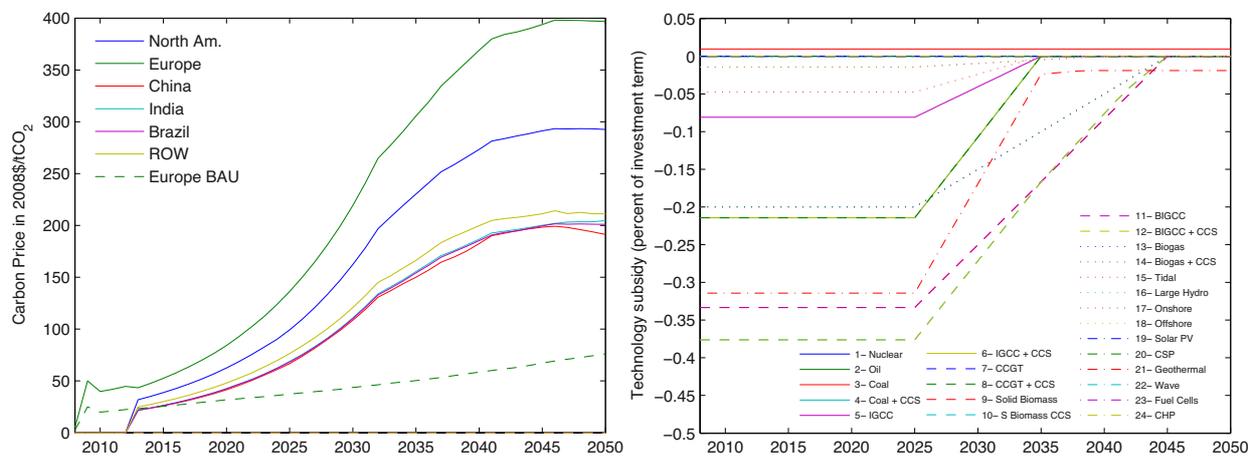}
		\end{center}
	\caption{(\emph{left}) Carbon price assumptions. Solid lines refer to the decarbonisation scenario while the dashed line refers to the baseline. (\emph{right}) Average technology subsidies for the decarbonisation scenario (defined as negative fractions of technology capital costs). The legend shows all FTT:Power technologies, however those not visible in the graph do not benefit from subsidies.}
	\label{fig:Figure2aSuppl}
\end{figure}

\textbf{Policy assumptions} The baseline (BAU) scenario assumes that current policies are maintained up to 2050, and this includes only carbon pricing in the EU through the Emissions Trading Scheme. For this, we used the projection given in 2008 dollars by the dashed line in fig.~\ref{fig:Figure2aSuppl}. In addition, feed-in tariffs exist in Japan, Germany, the UK and France for solar and wind technologies. These reduce the levelised cost of producing electricity \citep{Mercure2012} using these technologies to 5-15\% below the share weighed average cost of producing electricity in the grid, making these technologies competitive. No technology subsidies are present anywhere, however regulations to shut down existing nuclear power stations are in place in Germany and Japan. These are assumed to close down at the end of their lifetime and decrease exponentially from 2010 onwards. 

The 90\% decarbonisation scenario introduces carbon pricing in all regions of the World, with a summary for six regions given in extended data fig.~\ref{fig:Figure2aSuppl}, \ref{app:B} (averages across nations weighed by national emissions). Feed-in tariffs are introduced for solar and wind technologies in all regions, adjusted specifically to each regional situation. Technology subsidies are introduced for all low carbon technologies in all regions excluding those benefiting from feed-in tariffs \citep[the list of technologies used can be found in our previous work][]{Mercure2012,IEAProjCosts}. These range between 20 and 50\% of technology capital costs, adjusted to each specific regional situation. A global average is given in figure \ref{fig:Figure2aSuppl}. Complete information can be obtained from \cite{Mercure2014} and online on the 4CMR website at \burl{http://www.4cmr.group.cam.ac.uk/research/FTT/fttviewer}. 

\section{Additional data \label{app:B}}

\begin{figure}[h!]
		\begin{center}
			\includegraphics[width=1\columnwidth]{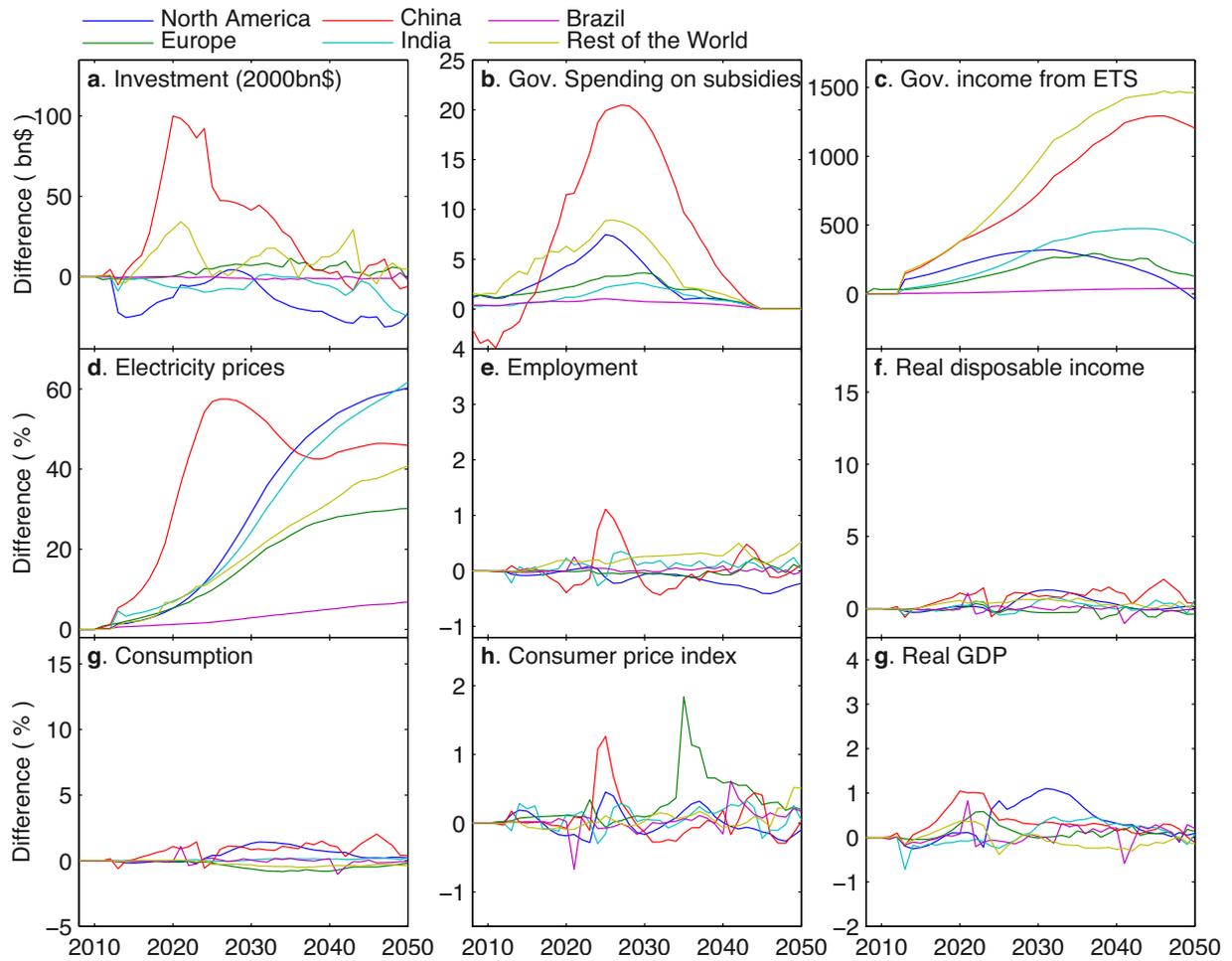}
		\end{center}
	\caption{Summary of economic changes due to climate policy resulting in the decarbonisation of the electricity sector in a scenario without revenue recycling. Values are expressed as changes from the baseline scenario in six aggregate region, in billions of nominal US dollars (\emph{a.} to \emph{c.}) or percent (\emph{d.} to \emph{h.}). No  losses of economic output are observed, and this is due to both a short run Keynesian effect and a long run productivity growth impact of investments in technology.}
	\label{fig:Figure1Suppl}
\end{figure}

\clearpage
\section*{Acknowledgements}

We acknowledge T. Barker for guidance, D. Crawford-Brown and Cambridge Econometrics for support, and M. Syddall for the online data explorer. This work was supported by the Three Guineas Trust (A. M. F.), Cambridge Econometrics (H. P. and U. C.), Conicyt (Comisi\'on Nacional de Investigaci\'on Cient\'ifica y Tecnol\'ogica, Gobierno de Chile) and the Ministerio de Energ\'ia, Gobierno de Chile (P. S.), the EU Seventh Framework Programme grant agreement n° 265170 `ERMITAGE' (N. E. and P. H) and the UK Engineering and Physical Sciences Research Council, fellowship number EP/K007254/1 (J.-F. M.). 

\section*{Author contributions}

J.-F. M. designed the research project, designed and built FTT:Power theoretically and computationally, integrated it to E3MG and performed the simulations. H.P. assessed the economic scenarios, coordinated the management and expansion of the E3MG model, and contributed to the economic theory discussion. U. C. contributed in creating the scenarios, modifying and maintaining E3MG as well as in the integration of FTT:Power. P. S. assisted FTT:Power work. A. M. F. ran the climate emulator simulations. N. E. and P. H. created the climate/carbon cycle emulators, facilitated their use and contributed to the text. All authors contributed to the conceptual content of the work. J.-F. M. wrote the manuscript.  

\bibliographystyle{spbasic}
\bibliography{../../../CamRefs}

\end{document}